# Icon Based Information Retrieval and Disease Identification in Agriculture

Namita Mittal
Dept. of Computer Engineering
MNIT, Jaipur

Basant Agarwal
Dept. of Computer Engineering
MNIT, Jaipur

Ajay Gupta
Dept. of Computer Engg.
MNIT, Jaipur,

Hemant Madhur
Dept. of Computer Engineering
MNIT, Jaipur

*Abstract*—**Recent developments in the ICT industry in past few decades has enabled the quick and easy access to the information available on the internet. But, digital literacy is the pre-requisite for its use. The main purpose of this paper is to provide an interface for digitally illiterate users, especially farmers to efficiently and effectively retrieve information through Internet. In addition, to enable the farmers to identify the disease in their crop, its cause and symptoms using digital image processing and pattern recognition instantly without waiting for an expert to visit the farms and identify the disease.**

**Keywords:** Iconic Interface, Image Processing, Pattern Recognition, Data Mining, Information Retrieval.

## I. INTRODUCTION

India currently has the largest illiterate population of any nation on earth. In present scenario, 25% people of India are language illiterate [9], digital illiteracy is far more than this. To facilitate the Internet opportunity to the common people, it is needed to represent the web content in their understandable and interactive form. The key problem involved here is the selection of an interactive form which is independent of any language. Iconic interface has emerged as a solution to this problem.

Farmers are facing problem in identifying diseases in crops, most of them are illiterate and unable to understand the reason, remedies etc. Most of the farmers are illiterate that's why they are not able to use internet for possible remedies of their infected crops. For this problem, iconic interface is proposed through which illiterate farmers can also select sequence of images for solution of their problem. This iconic interface builds a query on the basis of icons selected by the user. And for identifying diseases in crops, an automated system is proposed, this system is consist of the following stages image enhancement, HSI transformation, intensity adjustments using thresholding, segmentation, feature extraction and disease classification using neural network.

This paper discusses mainly two features. One with an iconic interface where the farmer can interact easily and in return system will return solution in native language. Another feature includes an image processing technique in which the farmer has to simply upload an image (with the assistance of an operator present at the Kisan help center) of the diseased crop and the processing result will show the name of the disease and possible remedial solution for infected crop.

In this system, user can select icons in an interactive way. An intermediate representation formed from the interaction result is fed to processing unit. The objective of this work is to cross the language barrier and make the Internet reachable to all groups of people. A user friendly iconic interface is proposed for the target user to achieve the goal. The iconic representation is powerful because it is independent of any languages. A proper sequence of iconic representation helps the user to follow a well-defined line of action.

Since long, disease identification has remained in the domain of experts examination and is mostly done manually. But the disease patterns vary in shape, size and textures which enables the development of an autonomous system for disease detection. The system has been developed for two diseases in the rice crop namely leaf blast and brown spot.

The remaining paper is organized as: Section II describes the related work, Section III explains the proposed approach, Section IV narrates results and Section V discusses the possible future work.

## II. RELATED WORK

One work related to iconic interface has developed a prototype for tourism industry [10]. Following works are related to the disease identification in agriculture using various image processing techniques. In [1], the classification of disease in corn using BP Neural Network is proposed, in which the RGB model is being converted into YCrCb model for segmentation. Another work corresponding to the grading of the plant diseases uses conversion of RGB model to HSI model and Ostu method for leaf region segmentation and Sobel operator for edge detection [2].One work related to the disease identification in the rice crop uses Support Vector Machine and extracts features based on shape and texture for classification. The System identifies namely three disease leaf blight, sheath blight and rice blast [3]. Another System aimed at identifying diseases in the rice crop uses unsupervised learning through SOM. The System uses entropy based thresholding and 8 connectivity methods for boundary detection [4].One of the related work corresponds to identification of Brown Spot disease in rice using BP Neural Network and color based features [5]. One such work aims at identifying the brown spot in rice crop uses K Means method for segmentation and BP neural network for classification of disease [6].

## III. PROPOSED APPROACH

1) Iconic Information Retrieval
2) Disease diagnosis using image processing techniques
   - Pre-processing Steps
   - RGB to HSI Conversion
   - Binary Saturated Mask Generation
   - Hue Image Masking With BSM

- Histogram Generation
- Thresholding
- Spot Detection Using Component Labeling
- Feature Extraction
- System Training

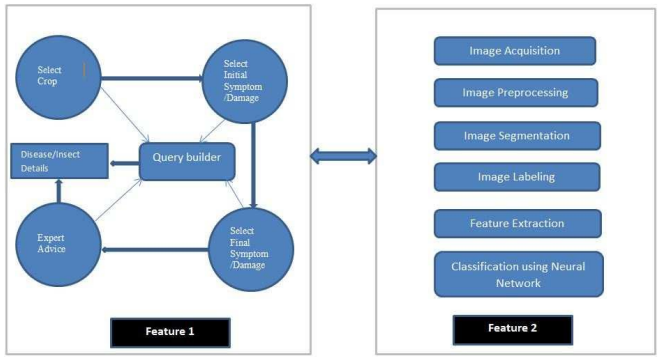

Fig. 1: Process Architecture

### A. Iconic Information Retrieval

The icons should be selected in such a way that they convey the information wholly as well as the user is not bombarded with too many icons to confuse him as well as preventing the redundancy in icon selection.

The icons are arranged in a hierarchical manner so the the query is generated with minimum number of keystrokes and also it is convenient for the user to understand and operate.

The query is generated from the selected icons and then fed to the system. The relevant information is retrieved from the database in accordance with icons selected for query generation.

### B. Disease diagnosis using image processing techniques

*1) Image Acquisition:* The samples of the infected rice leaves have been collected from google images.

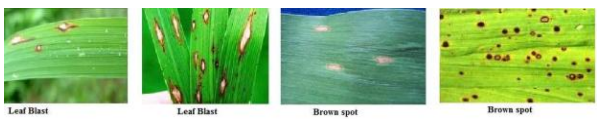

Fig. 2: Input Sample Image

*2) Pre-processing Steps:* As a pre-processing step the images used for training the system have been normalized to 200*200 size and are enhanced by increasing brightness and contrast.

*3) RGB to HSI Conversion:* The RGB images are then converted to the HSI model(more suitable for color description).The three components of the HSI model(H,S and I) are separated.

The RGB,CMY and other similar color models are not well suited for describing colors in terms that are practical for human interpretation. For example, one does not refer to the color of an automobile by giving the percentage of each of the primaries composing its color. When humans view a color object, we describe it by its hue, saturation and brightness(hue is the color attribute that describes a pure color, whereas saturation gives a measure of the degree to which a pure color is diluted by white color and brightness is a subjective descriptor that is practically impossible to measure). RGB model is suitable for color generation but HSI model is more suitable for color description [7].

The equation 1,2 and 3 shown below shows the conversion of translation rule for conversion of RGB model to HSI model.

$$H = ArcTan( Sqrt((3(G - R)/(R - G)(R - B))) \quad (1)$$

$$I = ((R + G + B)/3) \qquad (2)$$

$$S = I - Min(R, G, B)/3 \qquad (3)$$

The following figures 3,4 and 5 show the H,S and I images corresponding to the input images.

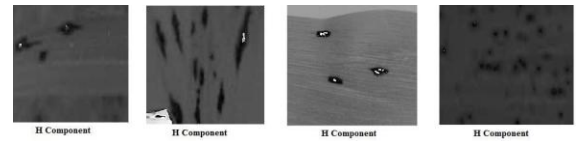

Fig. 3: H images

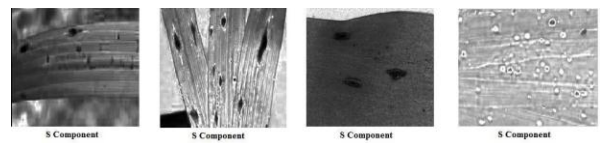

Fig. 4: S images

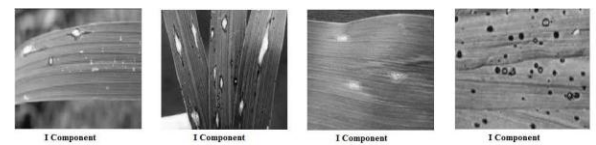

Fig. 5: I images

*4) Binary Saturated Mask Generation:* Binary saturated mask is generated by thresholding the saturation image. Suppose that the intensity histogram corresponding to an image, f(x,y) is composed of light objects on a dark background, in such a way that object and background pixels have intensity values grouped into two dominant modes. One obvious way to extract the objects from the background is to select a threshold, T, that separates these modes. Then, any point(x,y) in the image at which f(x,y) is greater than T is called an object point, otherwise, the point is called a background point [7].

*5) Hue Image Masking With BSM:* The hue image is then masked

with the binary saturated image(in order to isolate further regions of interest in the hue image).

Figure 6 shows the result of masking of Hue image with the Binary Saturated Mask.

*6) Histogram Generation:* A histogram of the image produced by the product of the hue image and the binary saturated image is then produced indicating the localization of the spots in a certain region. The histogram of a digital image with intensity levels in the range(0,L-1) is a discrete function

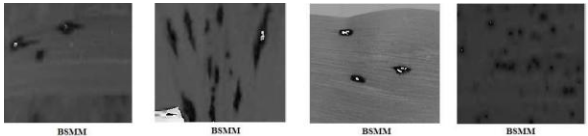

Fig. 6: Hue Image Masking With BSM

$$h(r_k) = n_k, \qquad (4)$$

where $r_k$ is the $k_{th}$ intensity value and $n_k$ is the number of pixels in the image with intensity $r_k$ [7]. Figure 7 shows the histogram generated for the Hue image masked with BSM.

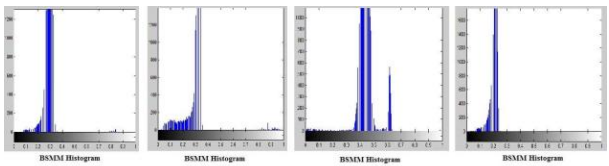

Fig. 7: Histogram

*7) Thresholding:* The masked image is then segmented using the thresholding. Figure 8 shows the segmented image for the corresponding input images.

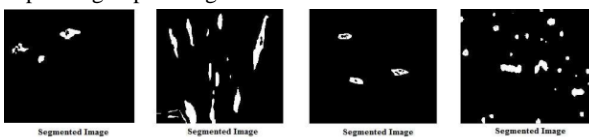

Fig. 8: Thresholding

*8) Spot Detection Using Component Labeling:* The spots boundaries are then detected using the image component labeling. Figure 9 shows a section of component labelling done.

*9) Feature Extraction:* The purpose of the feature extraction is to reduce the image data by measuring certain features or properties of each segmented region such as color,shape and texture. This phase mainly consists of two steps namely spot isolation and spot extraction. Often a segmented image consists of a number of spots. In order to extract features from individual spots, it is necessary to have an algorithm that identifies each spot.

To identify the spots, we label each spot with a unique integer and the largest integer label gives the number of spots in the image. Such identification algorithm is called component labeling. To recognize the images categorically, several features have to be identified for an image. The following features have been extracted from the images.

(a) Major Axis-The major axis points are the two points in an object where the object is more elongated and where the straight line drawn between these two points is the longest.Major axis points are calculated by all possible combinations of perimeter pixels where the line is the longest. The length of the major axis is given by:

$$MajorAxisLength = sqrt((x2-x1)^2 + (y2-y1)^2) \quad (5)$$

where (x1,y1) and (x2,y2) are the coordinates of the two end points of the major axis [8].

(b) Minor Axis-The minor axis is drawn perpendicular to the major axis where this line has the maximum length. Once the end points of the minor axis have been found, its length is given by the same equation as the major axis length. It is also called the object width [8].

(c) Bounding Box- The bounding box is defined by the smallest rectangle which encloses the object.
The minimum area of such a bounding box is given by:

$$BoundingBoxArea = majorAxisLength*minorAxisLength \quad (6)$$

(d) Eccentricity- Eccentricity is the ratio between the length of the short axis to the long axis.

$$Eccentricity = axisLengthShort/axisLengthLong \quad (7)$$

The value of eccentricity is between 0 and 1.Eccentricity is also called ellipticity with respect to minor axis and major axis of the ellipse. If the major axis gets longer, eccentricity gets higher [8].

(e) Perimeter-Perimeter is an important feature of an object. Contour based features which ignore the interior of a shape, depend on finding the perimeter or boundary points of the object [8]. The perimeter of an object is given by the integral as follows:

$$T = x(t) + y(t)dt \quad (8)$$

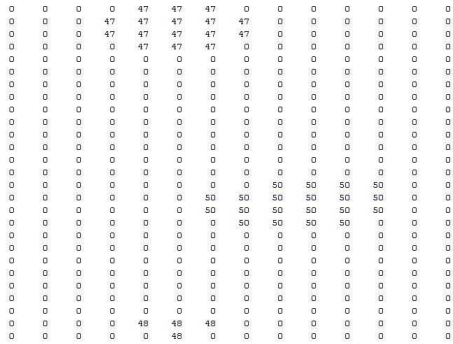

Fig. 9: Component Labeling

(f) Solidity- In simple terms density is mass per unit volume. But in two dimensional image objects this can be defined as the ratio between the area and convex area of the same object:

$$\text{Solidity} = \text{Area}/\text{ConvexArea} \qquad (9)$$

For a solid object or cell, this value is 1, while the value is lower for an object or cell having a rough perimeter or an object which has holes [8].

(g) Euler Number-Connected components and holes are important topological features and they are found out by the Euler number. The Euler number ( E ) is defined by the number of connected components (C ) and holes ( H ) :

$$\text{E} = \text{C} - \text{H} \qquad (10)$$

It is an important topological descriptor. This simple topological feature as said before is invariant to translation, rotation and scaling [8].

(h) Orientation- The overall direction of the shape. The direction of the major axis can be used as an (approximate) orientation for the object [8].

(i) Area-The number of pixels in the shape [8].

(j) Extent Measure-Extent measure also called as rectangularity ratio, has a value between 0 and 1.When this ratio has the value 1 then the shape is perfectly rectangular [8], is computed as

$$\text{Extent} = \text{SpotArea}/\text{BoundingBoxArea} \qquad (11)$$

(k) Diameter-Diameter of a spot is measured as

$$\text{Diameter} = \text{Sqrt}((4 * \text{Area}/\text{PI})) \qquad (12)$$

(l) Convex Hull-If the points are all on a line, the convex hull is the line segment joining the outermost two points. In the planar case, the convex hull is a convex polygon unless all points are on the same line. Similarly, in three dimensions the convex hull is in general the minimal convex polyhedron that contains all the points in the set.

The following table shows some of the features values of the images.

| Class | Area | Major axis | Minor axis | Eular No. |
|---|---|---|---|---|
| Leaf Blast | 4 | 3.44 | 1.69 | 1 |
| Leaf Blast | 5 | 4.25 | 1.94 | 1 |
| Brown Spot | 19 | 6.96 | 3.86 | 1 |
| Brown Spot | 11 | 4.49 | 3.28 | 1 |

*10) System Training:* This paper uses a feed forward neural networks with two hidden layers.

The system is trained for two classes (leaf blast and brown spot). The system is trained using 25 images for each kind of the disease. When a new image is fed to the system, it is classified under one of the two classes of the diseases. The system produces an output showing the name of the disease in the crop and its class.

## IV. RESULT

The icon based interface is shown in the figure 12 where the two broad categories for the selection are available. The selection of one of the icons will lead to the display of a sub-category of icons and so on. Finally a detailed description of the desired query will be presented in front of the user. System includes features like expert login, admin login, expert sign up, password security, contact us, feedback etc. Figure 10 and 11 shows the iconic interface and the categories under which the fields have been categorised respectively.

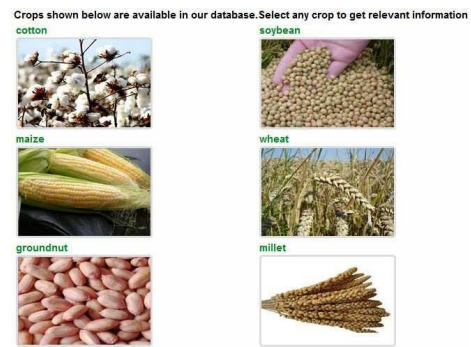

Fig. 10: Crop Icons

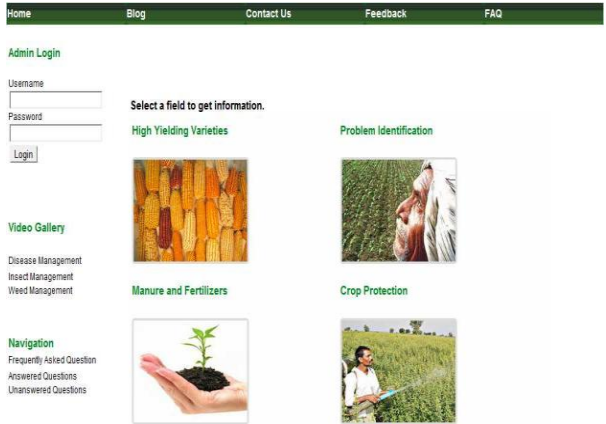

Fig. 11: Categories

**Additional features include-**

Blog-Those users who can read and write can directly interact with the experts through the blog.
Figure 12 shows the interface for the blog.

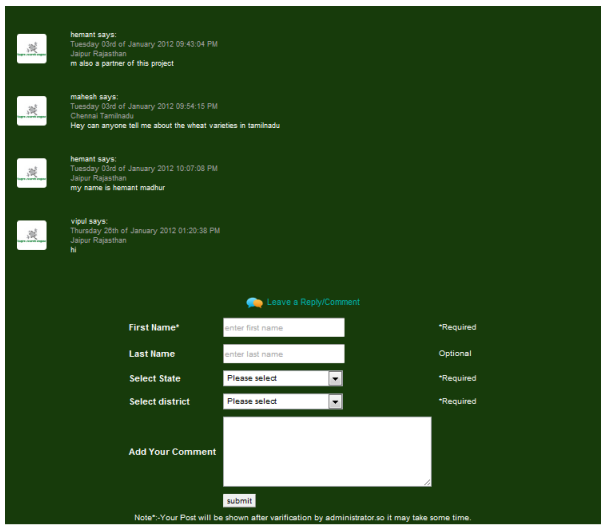

Fig. 12: Blog Screen

FAQ-Users can look for their queries directly in the frequently asked questions. Figure 13 shows the interface for the final detailed description shown to the user.

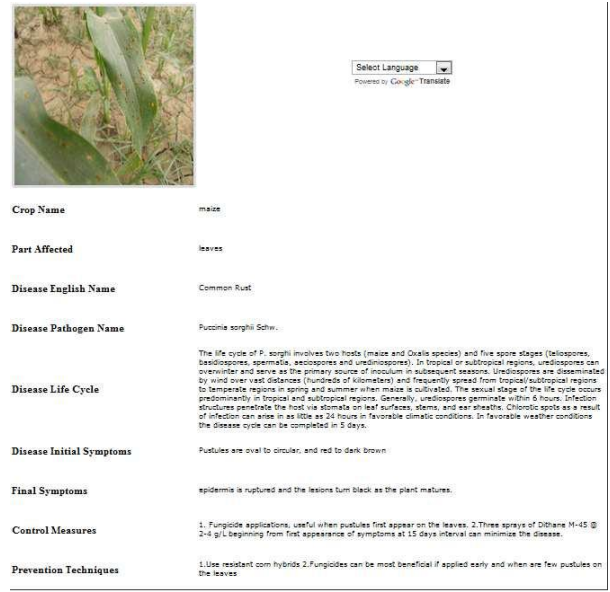

Fig. 13: Final Result Displayed to the User

Email-The user can also send an email to the experts regarding his query by simply click on a button.

To develop a full-fledged and ready to use system, some of issues have to be handled in icon selection, arrangement and image database handling. The advantage of using icons for query generation is that single icon is sufficient for the generation of a query but the disadvantage is that misinterpretation of the icon can lead to incorrect query formation. So, it is to be concluded that the use of iconic interface is very efficient as well as effective especially in the context of the naive users but its proper implementation is more difficult and complex task than it appears.

In the context of image processing section, back propagation neural network method has been used for disease classification in the rice crop for mainly two diseases. Morphological image descriptors have been used for image feature extraction. 25 images related to each category are used for training the system and 20 images of each category used for testing the system. Each spot on an image is taken as a separate input sample, thus increasing the number of input sample to 1008, and for testing each spot on an image is also taken as separate instance, and voting scheme is used to classifying the image in one of the two categories. The accuracy of the system based on the testing performed under this scheme comes out to be 89.23%.

V. Conclusion and Future work

This paper provides an interface model for digitally illiterate users, specially farmer to retrieve information through internet efficiently and effectively. It also enables farmers to identify the disease in their crop; its cause and symptoms using digital image processing and pattern recognition instantly without waiting for an expert to visit the farms and identify the disease. Experimental results prove the effectiveness of the proposed research.

The image processing section currently shows the results based its training by 25 images belonging to each category of the disease. The results can be further improved by increasing the number of images used for training of the system. The system currently works for a section of the crop and can be further extended for other crops as well as fruits and vegetables. Icon optimization by using synonyms and thesaurus is planned as the future work.